\documentstyle[12pt]{article}

\newcommand{\beq}{\begin{equation}}
\newcommand{\eeq}{\end	{equation}}

\newcommand{\beqar}{\begin{eqnarray}}
\newcommand{\eeqar}{\end  {eqnarray}}

\newcommand{\benum}{\begin{enumerate}}
\newcommand{\eenum}{\end  {enumerate}}

\newcommand{\bfig}{\begin{figure}}
\newcommand{\efig}{\end  {figure}}

\newcommand{\btab}{\begin{table}}
\newcommand{\etab}{\end  {table}}

\newcommand{\PRC}[1]{Phys. Rev. {\bf C{#1}}}
\newcommand{\PRD}[1]{Phys. Rev. {\bf D{#1}}}

\def \idag{\number\day .\space	%	Danish version of \today
\ifcase\month\or
januar\or februar\or march\or april\or maj\or juni\or juli\or
august\or september\or oktober\or november\or december\fi
, \number\year}

\begin{document}

\begin{titlepage}
\noindent
\vspace{7 mm}
%\hfill {\bf La Plata-Th96/19}\\
\hspace{1cm}
\begin{center}
{\bf
Density dependence of the MIT bag parameters
from the field theory of hadrons}
\vspace{20 mm}

{\sl R. Aguirre \footnote{Correspondence to: Dr. R. Aguirre,
E-mail: aguirre@venus.fisica.unlp.edu.ar .} and M. Schvellinger
\footnote{Fellow of the Consejo Nacional
de Investigaciones Cient\'{\i}ficas y T\'ecnicas, CONICET. \\
E-mail: martin@venus.fisica.unlp.edu.ar . }}

\vspace{3 mm}

Department of Physics, Universidad Nacional de La Plata, \\
C.C. 67 (1900) La Plata, Argentina.

\end{center}

\vspace{15 mm}

\begin{abstract}

A self-consistent description of the MIT bag parameters
as functions of the nuclear matter density is presented.
The subnuclear degrees of freedom are treated in the Quark-Meson Coupling
Model, considering the equilibrium conditions for the bag in the
nuclear medium.
The hadronic interaction is described in the framework of
the quantum field theory of hadrons
through several models. We have obtained the behavior
of the bag radius and the bag parameters $B$ and $z_0$, taking
their derivatives with respect to the mean field value of the scalar
meson as free parameters. A discussion on the
variation range of these derivatives is
given.

\end{abstract}


\end{titlepage}

It is well known that Quantum Chromodynamics (QCD)
is the most accepted candidate for a
theory of strong interactions. Although QCD has been successful
in describing the high energy domain, its application seems difficult
in the energy range
associated with nuclear phenomena. Attempts to
sort this difficulty have been carried out by using effective models,
among them the so-called bag models are widely used \cite{CHODOS,THOMAS1}.
They provide an acceptable description of the free hadronic properties.
The energy-momentum conservation of an isolated bag is imposed through
the so-called non-linear boundary condition \cite{CHODOS,THOMAS1}.

The Quark-Meson Coupling Model (QMC) has been developed
since the early work of Guichon \cite{GUICHON}.
In this model the nucleons are described
as non-overlapping MIT bags
confining the quarks inside them. The quarks interact
through the exchange of
scalar ($\sigma$) and vector (${\omega}_{\mu}$) mesons.
Several refinements and applications
of this model have been done \cite{SAITO1}--\cite{CMC}, for instance
the QMC model has been used to describe nuclear and neutronic matter
\cite{SAITO1,SAITOHEP}, as well as finite nuclei \cite{SAITOHEP,HEPS}.

On the other hand, the nuclear phenomena has been
treated in the context of quantum field theory
by using structureless
hadronic fields as the effective degrees of freedom. This
subject reached interest since the pionnering work of Walecka
\cite{W0}. Different aspects of the phenomenology of nuclear
matter and finite nuclei have been studied in the framework of the
so-called Quantum Hadrodynamics (QHD) with successful results
\cite{WALECKA}. The original Walecka model has been
extended with additional polynomic potentials \cite{PP} and alternative
non-polynomic interactions in the scalar channel \cite{ZM}-\cite{FELDMEIER}.

If the validity of the equation of state , effective mass, etc.,
derived from an effective hadronic lagrangian is assumed, one can ask
about the hadronic substructure consistent with those properties.
We are interested in the implications that the QHD description
have for a picture which deals with subnuclear degrees of freedom.
A clear relationship between the QMC model and
the Walecka model has been stablished by Saito and Thomas \cite{SAITO1} .
In the present work we have
studied the reliability of a coherent description of in-medium nucleon
properties in terms of QHD and QMC.

We briefly recall some basic features of the QMC model \cite{GUICHON,SAITO1}.
If isospin symmetry breaking is neglected the meson fields $\sigma (x)$
and $\omega_{\mu} (x)$ are
sufficient to describe the problem. Inside the bag the equation of
motion for quarks of mass $m_q$ is given by
\beq
(i {\not\!{\partial}} - m_q) {\Psi}_q(x) = [-g^q_{\sigma} \sigma (x) +
g^q_{\omega} {\not\!{\omega (x)}}] {\Psi}_q(x) ,
\eeq
where $g^q_{\sigma}$ and $g^q_{\omega}$ are the quark-meson coupling
constants associated with the $\sigma$ and $\omega_{\mu}$ fields, respectively.
In the Mean Field Approximation
(MFA) the meson fields are replaced by their
mean values, which become constants in infinite nuclear matter, i.e.
$\sigma=\bar{\sigma}$ and $\omega_{\mu}=\bar{\omega} \delta_{\mu 0}$.

The normalized quark wave function for a spherical bag of radius $R$ is
\beq
{\Psi}_q({\vec{r}},t) = {\cal{N}} e^{-i {\epsilon}_q t/R}
\times \left( \begin{array}{c} {j_0(y r/R)} \\	i {\beta}_q
{\vec{\sigma}} \cdot $\^{r}$ j_1(y r/R)  \end{array} \right)
{{{\chi}_q}\over{\sqrt{4 \pi}}} ,
\eeq
where $r=|{\vec{r}}|$,
${\chi}_q$ is the quark spinor and the normalization constant
${\cal{N}}$ is given in \cite{GUICHON}.

We have introduced the effective quark mass
$m^*_q = m_q -g^q_{\sigma} {\bar{\sigma}}$ and the
effective energy eigenvalue
${\epsilon}_q = \Omega/R+ g^q_{\omega} {\bar{\omega}}$,
where ${\Omega} = \sqrt{y^2 + (R m^*_q)^2}$. The $y$ variable is
fixed by the boundary condition at the bag surface
$j_0(y) = {\beta}_q j_1(y)$ as in reference \cite{CHODOS} and
${\beta}_q = \sqrt{({\Omega} - R m^*_q)/({\Omega} + Rm^*_q)}$.

The bag energy is given by
\beq
E_b = {{3 {\Omega} - z_0}\over{R}} + {{4}\over{3}} \pi B R^3
\label{BAGENERGY}
\eeq
where $B$ is the energy per unit of volume and $z_0$ takes into account the
zero point energy of the bag.
The nucleon mass is defined by including the correction due to the
spurious center of mass motion \cite{CMC}
\beq
M^*_b = \sqrt{E_b^2 - 3 {(y/R)}^2} .
\label{BAGMASS}
\eeq

It is usual to determine the bag parameters at
zero baryon density to reproduce the experimental nucleon mass
$M_b = 939$ MeV. Simultaneously it is required the equilibrium condition
for the bag $d M_b( {\bar{\sigma}} )/d R = 0$.
In \cite{SAITOHEP} the bag parameters $B$ and $z_0$ are constants,
although it was found a relative change of $1 \%$ in the radius
at the saturation density (${\rho}=0.15$ fm$^{-3}$)
as compared with its value at zero baryon density.

In the normal QMC model the bags interact  by the same mechanism
that couples mesons to quarks inside a bag. In the present work
we have proposed a description of the hadronic interaction by
a set of different effective models commonly used in QHD.
In all the cases considered we have used a linear coupling between the nucleon
field $\Psi (x)$ and the vector-meson field, with strength $g_{\omega}$.
However the  models differ in the interaction term
${\bar{\Psi}}(x) V_{N \sigma}{\Psi} (x)$,
between the nucleon and the scalar meson field.
In Table I we show the explicit form of $V_{N \sigma}$
for these four models.
The coupling constants $g_{\sigma}$ and $g_{\omega}$ have been fixed
to reproduce the binding energy per nucleon, $E_b = 16$ MeV, and the
nuclear saturation density, ${\rho} = 0.15$ fm$^{-3}$, in the MFA.
The isothermal compressibility $\kappa = 9 {\rho}\,( \partial P_h/\partial {\rho}_B)_T$
is a usefull quantity in order to analyze the fitness of hadronic
models, since its value at the nuclear saturation  density has been
well determined to range between $100-300$ MeV \cite{BLAIZOT}.
The isothermal compressibility evaluated in the MFA
is also shown in Table I \cite{FELDMEIER}.

The Euler-Lagrange equations for nucleons and mesons in the MFA
are given by
\beq
(i {\not\!{\partial}} - M_N) {\Psi}(x) =
(g_{\omega} {\bar{\omega}} {\gamma}_0 - V_{N \sigma}) {\Psi}(x) ,
\eeq
\beq
m^2_{\sigma} {\bar{\sigma}} =
\frac{d V_{N \sigma}}{d \sigma}(\sigma=\bar{\sigma})
\, \, \, {\rho}_s ,
\label{HADRONSELF}
\eeq
\beq
m^2_{\omega} {\bar{\omega}} = g_{\omega} {\rho} ,
\eeq
where ${\rho}_s = < {\bar{\Psi}}(x) {\Psi}(x) >$, ${\rho}$ is the
baryon density; $M_N$, $m_{\sigma}$ and $m_{\omega}$ are the masses
of the free nucleon, the scalar  and vector mesons,
respectively.

At zero temperature we have
\beq
{\rho}={{2 k^3_F}\over{3 {\pi}^2}} ,
\eeq
and
\beq
{\rho}_s= {{4}\over{(2 \pi )^3}} \int d^3{\vec{k}} \Theta (k_F-|\vec{k}|)
{{M^*_N}\over{\sqrt{{M^*_N}^2+{\vec{k}}^2}}} ,
\eeq
where $k_F$ is the nucleon Fermi momentum.
The effective nucleon mass $M^*_N$
and the energy spectrum $\epsilon (k)$ are given by
\beq
M^*_N = M_N - V_{N \sigma},
\label{HADRONMASS}
\eeq
and
\beq
\epsilon (k) = \sqrt{{M^*_N}^2+k^2} + g_{\omega} {\bar{\omega}} .
\eeq
We have used different notations for the nucleon mass entering
in QHD ($M^*_N$), and the nucleon mass generated
by the bag model ($M^*_b$).

Equation (\ref{HADRONSELF}) is a self-consistent definition for the mean
field value $\bar{\sigma}$, indeed from this equation
we see that the derivative  $d M^*_N/d\bar{\sigma}$ determines
the dynamics of the scalar field.
The QMC and QHD descriptions produce coherent results if
the following equation is fulfilled
\beq
M^*_N( \sigma ) = M^*_b( \sigma ) ,
\label{IGUALMASS}
\eeq
together with $g_{\sigma}=3 g^q_{\sigma}$, $g_{\omega}=3 g^q_{\omega}$,
\cite{SAITO1}.

The stability of the bag in the nuclear medium with respect to volume
changes is imposed by
\beq
P_b( \sigma )=P_h( \sigma ) ,
\label{IGUALPRESS}
\eeq
where $P_b( \sigma )$ is the internal pressure generated by the quark
dynamics and $P_h( \sigma )$ is the external hadronic pressure.

The bag pressure can be written as
\begin{eqnarray}
P_b  &=&-\frac{E_b}{4 \pi R^2  M^*_b}\left(-\frac{E_b}{R}+ \frac{16}{3}
\pi R^2 B+ \frac{3 m^{* 2}_q}{\Omega} \right)-\frac{3 y^2}
{4 \pi R^5 M^*_b},
\label{BAGPRESS}
\end{eqnarray}
and the pressure for uniform nuclear matter is given by
$P_h = - {{1}\over{3}} T^{ii}$,
where $T^{ii}$ is the trace over the spatial components of the
energy-momentum tensor
\beq
P_h = \frac{2 M^{*3}_N}{\pi^2} \left[ \frac{k_F}{M^*_N}
\left(\frac {5 \epsilon_F}{24} - \frac{\epsilon_F^3}{12 M^{*2}_N}
\right) - \frac{M^*_N}{8} ln \left( \frac{\epsilon_F+k_F}{M^*_N}
\right) \right].
\label{HADRONPRESS}
\eeq

The equation (\ref{IGUALPRESS}) is a statistical equilibrium condition
on the bag surface which ensures a direct relation between nuclear
matter bulk properties and the stability of the confining volume.

The equations (\ref{IGUALMASS}) and (\ref{IGUALPRESS})
can be used to obtain
the bag parameters as functions of the density. However,
they are not enough to determine all the
functions involved. Henceforth we have included the additional equations
obtained from the derivatives of the boundary condition and of the
equations (\ref{IGUALMASS}) and (\ref{IGUALPRESS}), taking
$y, m^*_q, R, z_0$ and $B$ as independent functions. In fact we
have considered $\lambda = d B/ d \bar{\sigma}$ and
$\mu = d z_0/d \bar{\sigma}$ as constants.
This procedure is equivalent to take only the linear contributions
in the expansion of the functional relations
(\ref{IGUALMASS}) and (\ref{IGUALPRESS}).

After the hadronic coupling constants have been adjusted to reproduce
the saturation conditions for nuclear matter, equation (\ref{HADRONSELF})
can be used to obtain the solution $\bar{\sigma}$
for each baryon density. Equations (\ref{BAGENERGY}), (\ref{BAGMASS}),
(\ref{HADRONMASS}) and (\ref{IGUALMASS})
can be used together with eqs. (\ref{IGUALPRESS}), (\ref{BAGPRESS}) and
(\ref{HADRONPRESS}) to determine $B$ and $z_0$ when the
value of the in-medium bag radius is provided. The last quantity is obtained
by solving self-consistently the additional equations
for fixed values of the parameters $\lambda$ and $\mu$.
To search for appropriate values of $\lambda$ and $\mu$
we have evaluated $R, B$ and $z_0$ at zero baryon density as a function
of $( \lambda , \mu )$ using the model 1. A drastic change in the behavior
of the quantities considered is found when $\mu$
goes from negative to positive values.

The bag radius $R$ at the saturation density has been studied
as a function of $\lambda$.
For the hadronic models considered here and taking the bag radius at zero
baryon density $R_0 = 0.6$ fm we have found that $R/R_0 > 1$
only for a restricted range of $\mu$.
For the following discussion we have taken two sets of values;
set I ($\lambda=-5.28$ fm$^{-3}$, $\mu=-0.50$ fm) and
set II ($\lambda=0, \mu=1.6$ fm).

The bag radius as a function of the density is shown in fig. 1,
the quark mass at zero baryon density has been fixed at $m_q=10$ MeV.
It can be seen that set I gives an asymptotical constant
bag radius for every model,
whose values do not depend on the details of the
interaction and it is diminished as compared with its vacuum value.
The set II provides a
model dependent radius at high densities, in this case models 1 and 2
predict a breaking-down of the bag picture.
Models 3 and 4 have a
stable behavior for all the densities considered here, even when set II
is used.

In figs. 2 and 3 we present the density dependence of $B^{1/4}$ and of
$z_0$, respectively. Models 1 and 2 exhibit an opposite behavior
for $B^{1/4}$, when set I or set II are used.

For model 1 and  set I $B$
takes negative values as the baryon density is sufficiently increased,
thus the bag bulk energy must decrease with increasing volume. The increment
of the kinetic energy compensates this fact, giving a slowly decreasing
total bag energy and a stable bag radius (see fig. 1).
On the other hand, $B$ grows drastically at high densities for set II,
a small volume increment gives rise to a large
increment in the bulk energy. Therefore in order to get a slowly
decreasing $M^*_b$, the bag radius
$R$ must decrease at the same rate as $B^{1/3}$ grows.
When $R$ approaches to zero a subtle
cancellation among the
quark kinetic energy, the zero point motion parameter and the center
of mass correction takes place.
The raising of $z_0$ (see fig. 3) is not sufficient to reach the
dynamical equilibrium and hence the system reduces its volume as far as
possible.

The steep behavior of models 1 and 2 as compared with models 3 and 4 is due
to the fact that the first mentioned models give a stiff equation
of state and a fast decrease for the effective nucleon mass.

Our results can be compared with those obtained by Jin and Jennings
\cite{XJ1}. In their work a phenomenological description of $B$ as a function
of the density is given in terms of a set of two free
parameters with no direct dynamical interpretation, namely
$g_{\sigma}^B, \delta$	and $B_0$ for model I of \cite{XJ1}.
Starting from the usual approach in the QMC model, in ref. \cite{XJ1}
the constant $B$ is replaced by a function of the baryon density.
Within this framework they found a monotonous density dependence of the bag
parameter $B$.
For $\delta = 4$ and $g_{\sigma}^B = 1$, the ratio $B/B_0$
decreases $80 \%$ at the nuclear saturation density and the corresponding
in-medium bag radius increases $60 \%$.
In our approach we have obtained an increase of $12 \%$ for the
bag radius (fig. 1) and a corresponding decrease of $19 \%$ in $B$ (fig. 2),
when model 4 and the set II of parameters are used.

In this work we have studied the coherence of the
QHD and QMC descriptions by using the
equilibrium conditions for the bag in nuclear matter.
Thus we have stablished a link between these frameworks which is able
to explain the nucleon substructure changes taking into
account the nuclear matter equation of state, the nucleon effective mass, etc.

We have evaluated the density dependence of the bag parameters and the
bag radius. In our model we use two dynamical quantities, i.e. the derivatives
$dB/d \bar{\sigma}$ and $dz_0/d \bar{\sigma}$
as free parameters and we have explored their possible variation range. We have
found two different dynamical regimes for these parameters.

The inclusion of thermal effects and the study of the EMC effect
in this framework will be reported in  future works \cite{HEPQMC}.

\newpage

\begin{quote}

\centering Table I.

\begin{tabular}{ccccc}	     \hline\hline
$Model$  &$V_{N \sigma}$				 & $g_{\sigma}$  & $g_{\omega}$ & $\kappa$ \\ \hline
	 &						 &		 &		& [MeV] \\ \hline
$  1 $	&$  g_{\sigma} {\sigma}$			&    11.04	&   13.74      &  554  \\
$  2 $	&$M_N g_{\sigma}tanh(g_{\sigma} \sigma /M_N)$	&     9.15	&   10.52      &  410  \\
$  3 $	&$M_N [1-exp(g_{\sigma} \sigma /M_N)]$		&     8.34	&    8.19      &  267  \\
$  4 $	&$g_{\sigma}{\sigma}/(1+ g_{\sigma}\sigma /M_N)$&     7.84	&    6.67      &  224  \\   \hline\hline
\end{tabular}

\end{quote}

Table I: Nucleon-scalar meson interaction terms,
coupling constants and isothermal
compressibility $\kappa$, for several effective hadronic models
used in this work.

\newpage

\vfill
\eject
\centerline{Figure Captions}
\begin{description}

\item [Figure 1:]
The bag radius $R$
as a function of the relative baryon density ${\rho}/{\rho}_0$,
where $\rho_0$ is the nuclear matter density at saturation,
for hadronic models indicated as M 1 (Walecka model), M 2,
M 3, and M 4 (Zimanyi-Moszkowski model).
Full and dashed lines correspond to set I and set
II of parameters, respectively. The quark mass at zero baryon density has been
taken as $m_q=10$ MeV.

\item [Figure 2:]
$B^{1/4}$ in [MeV]
as a function of the relative baryon density ${\rho}/{\rho}_0$.
The conventions and parameters used are the same as in Fig.1.

\item [Figure 3:]
$z_0$
as a function of the relative baryon density ${\rho}/{\rho}_0$.
The conventions and parameters used are the same as in Fig.1.

\end{description}
\vfill
\eject

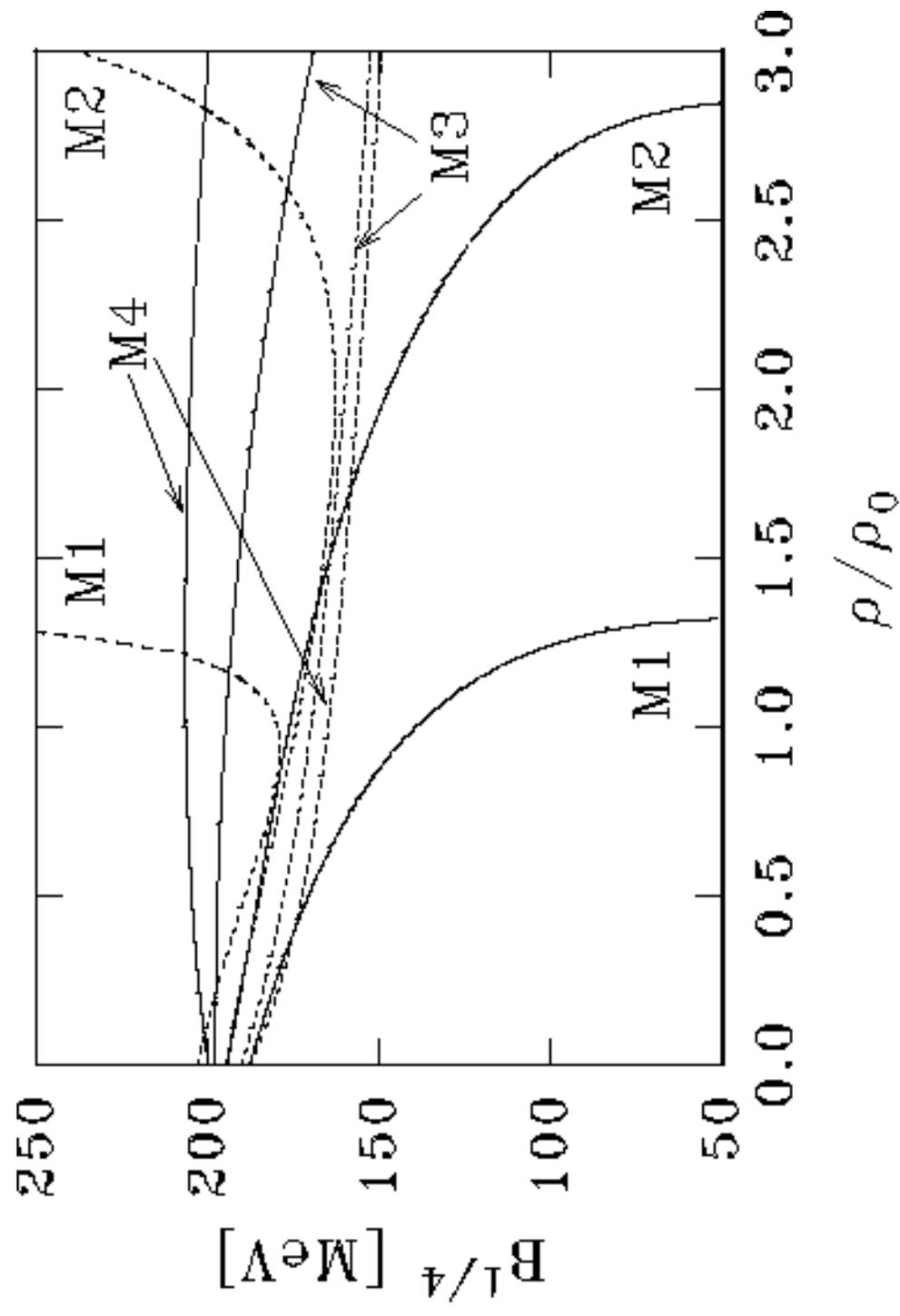
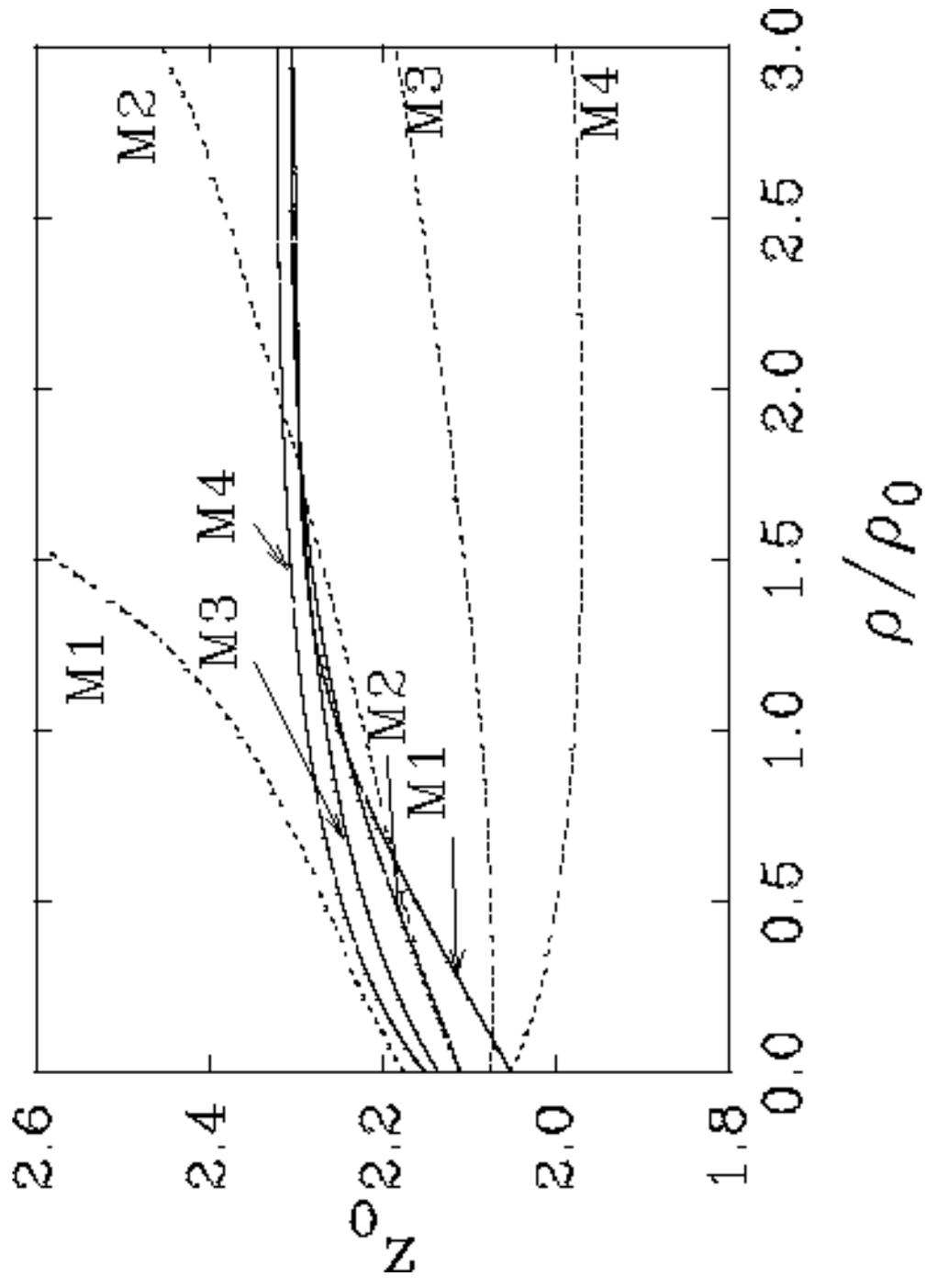
\end{document}